\documentclass[preprint,11pt, a4paper]{elsarticle}

\usepackage[
  top=1in,
  bottom=1in,
  left=1in,
  right=1in
]{geometry}

\usepackage{amssymb}
\usepackage{hyperref}

\usepackage{float}
\restylefloat{table}

\usepackage[utf8]{inputenc}
\usepackage[T1]{fontenc}
\usepackage{textcomp}
\usepackage{soul}
\usepackage{graphicx,subfig}
\usepackage{amsmath,array,amssymb,xparse,upgreek,bm}
\usepackage{tikz,mathpazo}
\usetikzlibrary{shapes.geometric,arrows,positioning,fit,backgrounds,calc}
\usepackage{hyperref}
\usepackage{tcolorbox}
\usepackage{booktabs}
\usepackage{multirow}
\DeclareMathOperator*{\argmin}{arg\,min}
\usepackage[ruled]{algorithm2e}
\usepackage{empheq}
\usepackage{tablefootnote}
\usepackage{array}
\usepackage{xcolor}

\newcommand{\symYes}{\tikz[baseline=-0.5ex]\fill (0,0) circle (0.62ex);}
\newcommand{\symNo}{\tikz[baseline=-0.5ex]\draw[line width=0.4pt] (0,0) circle (0.62ex);}
\newcommand{\symPart}{\tikz[baseline=-0.5ex]{\draw[line width=0.4pt] (0,0) circle (0.62ex); \fill (0,0) -- (90:0.62ex) arc (90:270:0.62ex) -- cycle;}}

\journal{SoftwareX}

\begin{document}
\renewcommand{\labelenumii}{\arabic{enumi}.\arabic{enumii}}

\begin{frontmatter}



\title{pyALDIC: A Python Implementation of Augmented Lagrangian Digital Image Correlation with a GUI, Adaptive Meshing, and Mask-Aware Subset Splitting}

\author[utae]{Zixiang Tong}

\author[utae,tmi]{Jin Yang\corref{cor1}}
\ead{jin.yang@austin.utexas.edu}

\cortext[cor1]{Corresponding author.}

\address[utae]{Department of Aerospace Engineering and Engineering Mechanics, The University of Texas at Austin, Austin, TX 78712, USA}
\address[tmi]{Texas Materials Institute, The University of Texas at Austin, Austin, TX 78712, USA}

\begin{abstract}
pyALDIC is an open-source Python implementation of augmented Lagrangian digital image correlation (AL-DIC) for full-field displacement and strain measurement. The software combines a graphical user interface with a scriptable Python API and supports adaptive quadtree meshing, mask-aware subset splitting near cracks and holes, and selectable Local DIC and AL-DIC solver modes. Numba acceleration enables efficient analysis, while automated tests, documentation, and reproducible examples support reliable use across Windows, macOS, and Linux. Verification cases include synthetic displacement fields, rigid-body motion, Mode-I cracking, adaptive refinement, and experimental uniaxial tension. pyALDIC is distributed through PyPI, GitHub, and Zenodo under a BSD-3-Clause license for reproducibility. pyALDIC is openly available at \url{https://github.com/zachtong/pyALDIC}.
\end{abstract}


\end{frontmatter}


\newpage
\section*{Required metadata}

\section*{Current code version}

\begin{table}[H]
\centering
\begin{tabular}{|l|>{\raggedright\arraybackslash}p{5cm}|>{\raggedright\arraybackslash}p{6.3cm}|}
\hline
\textbf{Nr.} & \textbf{Code metadata description} & \textbf{Please fill in this column} \\
\hline
C1 & Current code version & v0.6.0 \\
\hline
C2 & Permanent link to code / repository used for this code version & (i) \url{https://github.com/zachtong/pyALDIC}; (ii) \url{https://zenodo.org/records/21131340} \\ 
\hline
C3 & Code Ocean compute capsule & --- \\
\hline
C4 & Legal Code License & BSD 3-Clause \\
\hline
C5 & Code versioning system used & git \\
\hline
C6 & Software code languages, tools, and services used & Python (version $\geq 3.10$); PySide6; NumPy; SciPy; OpenCV; scikit-image; Numba; imageio; matplotlib; pytest \\
\hline
C7 & Compilation requirements, operating environments \& dependencies & OS-independent (Windows, macOS, Linux). Install with \texttt{pip install al-dic}, or editable install for development with \texttt{pip install -e ".[dev]"}. No compilation required; Numba JIT-compiles solver kernels at first invocation. \\
\hline
C8 & If available, link to developer documentation / manual & \url{https://github.com/zachtong/pyALDIC/tree/main/docs/user-guide} \\
\hline
C9 & Support email for questions & \href{mailto:zachtong@utexas.edu}{zachtong@utexas.edu} \\
\hline
\end{tabular}
\caption{Code metadata for pyALDIC.} 
\label{tab:code_metadata}
\end{table}

\newpage
\section*{Current executable software version}

\begin{table}[h!]
\centering
\begin{tabular}{|l|>{\raggedright\arraybackslash}p{5cm}|>{\raggedright\arraybackslash}p{6.3cm}|}
\hline
\textbf{Nr.} & \textbf{(Executable) software metadata description} & \textbf{Please fill in this column} \\
\hline
S1 & Current software version & v0.6.0 \\
\hline
S2 & Permanent link to executables of this version & (i) \url{https://pypi.org/project/al-dic/0.6.0/}; (ii) \url{https://zenodo.org/records/21131340} \\
\hline
S3 & Legal Software License & BSD 3-Clause \\
\hline
S4 & Computing platforms / Operating Systems & OS-independent (Windows, macOS, Linux) \\
\hline
S5 & Installation requirements \& dependencies & Python (version $\geq 3.10$); PySide6; NumPy; SciPy; OpenCV; scikit-image; Numba; imageio; matplotlib (installed automatically by \texttt{pip install al-dic}) \\
\hline
S6 & If available, link to user manual & \url{https://github.com/zachtong/pyALDIC/tree/main/docs/user-guide} \\
\hline
S7 & Support email for questions & \href{mailto:zachtong@utexas.edu}{zachtong@utexas.edu} \\
\hline
\end{tabular}
\caption{Executable software metadata for pyALDIC v0.6.0, installable from PyPI as \texttt{al-dic}.}
\label{tab:software_metadata}
\end{table}

\newpage
\section{Motivation and significance}
\label{sec:intro}

Digital Image Correlation (DIC) is widely used in experimental mechanics for measuring full-field deformations and characterizing the mechanical behavior of materials~\cite{sutton2009,pan2018review,idics_gpg_2018}. In a typical DIC experiment, a sequence of images is acquired from the surface of a specimen that is usually painted with a random speckle pattern. By comparing the undeformed reference image with one or multiple deformed images, DIC tracks the motion of material points within a user-defined region of interest (ROI) and a defined DIC mesh. The resulting nodal displacement fields can be subsequently differentiated to obtain strain fields. Among the available DIC post-processing formulations, local subset-based image tracking methods (e.g., the inverse-compositional Gauss--Newton (IC-GN) algorithm \cite{baker2004lucas,pan2013fast}) are widely used because of their computational efficiency. However, their performance degrades under several challenging conditions that arise frequently in experimental practice, including low signal-to-noise images \cite{yang2019combining}, large displacement gradients \cite{gupta2024estimation}, and {discontinuous} kinematic fields such as associated with cracks \cite{poissant2010novel,zhu2026pf}, shear bands \cite{chen2017active,ni2025revisiting,ran2026comparative}, and interfacial debonding or sliding \cite{rubino2019full}. In particular, a correlation subset that straddles a displacement discontinuity may contain material regions undergoing distinct motions and therefore has no single coherent match in the deformed image. \\

To address these challenges, we previously introduced a hybrid local--global DIC image tracking formulation, called \textbf{Augmented Lagrangian DIC (AL-DIC)}~\cite{yang2019augmented}, in which independent local correlation problems are coupled to a global solver through the alternating direction method of multipliers (ADMM)~\cite{boyd2011distributed}. The global step enforces kinematic compatibility, suppresses noise, and reduces isolated erroneous solutions that may arise from purely local correlation~\cite{yang2019combining}. The performance of AL-DIC has been evaluated using the DIC Challenge 2.0 datasets and compared with that of other commonly used commercial and academic DIC codes \cite{dicchallenge2exme}. In addition, the AL-DIC formulation has subsequently been extended through adaptive spatial discretization to enhance its computational efficiency  \cite{yang2021fast,yang2018fast,yang2022staq,leu2026machine}. This framework has also been extended to three-dimensional digital volume correlation (DVC) \cite{yang2020augmented} and 3D stereo-DIC for measuring out-of-plane deformation and deformation on non-planar surfaces~\cite{tong2025stereo3d}. \\

\begin{table}[h!]
\small
\centering
\resizebox{1 \textwidth}{!}{
\begin{tabular}{p{2.6cm}p{0.7cm}p{1.4cm}>{\centering\arraybackslash}p{1.1cm}>{\centering\arraybackslash}p{1.3cm}>{\centering\arraybackslash}p{1.3cm}>{\centering\arraybackslash}p{1.3cm}>{\centering\arraybackslash}p{1.4cm}}
\toprule
Name & Refs & Language & GUI & Local method & Global method & Adaptive mesh & Mask-aware subsets \\
\midrule
Ncorr            & \cite{blaber2015ncorr}      & MATLAB       & \symYes  & \symYes & \symNo  & \symNo  & \symPart   \\
DICe             & \cite{turner2015dice}       & C\texttt{++} & \symYes & \symYes & \symYes & \symNo  & \symPart   \\
$\mu$DIC         & \cite{olufsen2020mudic}     & Python       & \symNo   & \symNo  & \symYes & \symNo  & \symNo   \\
OpenCorr         & \cite{li7088402opencorr}             & C\texttt{++} & \symYes  & \symYes & \symNo  & \symNo  & \symNo   \\
iCorrVision-2D   & \cite{icorrvision}          & Python       & \symYes  & \symYes & \symNo  & \symNo  & \symNo   \\
UFreckles        & \cite{rethore2018ufreckles} & MATLAB       & \symYes  & \symNo  & \symYes & \symYes & \symNo   \\
YaDICs           & \cite{yadics}               & C\texttt{++} & \symNo   & \symYes & \symYes & \symNo  & \symNo   \\
PReDIC           & \cite{predic}               & Python       & \symNo   & \symYes & \symNo  & \symNo  & \symNo   \\
ALDIC & \cite{yang2019augmented}   & MATLAB       & \symNo   & \symYes & \symYes & \symNo  & \symNo   \\
\textbf{pyALDIC  (this work)} & \cite{pyaldic_zenodo} & \textbf{Python} & \symYes & \symYes & \symYes & \symYes & \symYes   \\
\bottomrule
\end{tabular}}

\vspace{4pt}
{\footnotesize \symYes~yes\quad \symNo~no\quad \symPart~partial.\par}
\caption{ Comparison of representative open-source DIC software packages.
``Local method'' and ``Global method'' indicate whether a package provides a local subset-based correlation formulation, a global finite-element or regularized formulation, or both.
AL-DIC couples local and global formulations through an augmented-Lagrangian framework \cite{yang2019augmented}.
``Adaptive mesh'' denotes spatial refinement of correlation points or finite-element nodes.
``Mask-aware subsets'' denotes treatment of correlation subsets that intersect mask boundaries associated with cracks, holes, or ROI boundaries.
``Partial'' indicates that masked pixels can be excluded from the correlation calculation, whereas ``Yes'' indicates connected-component-based subset splitting that retains only the valid region containing the subset center.  } \label{tab:dic_codes}
\end{table}

To date, AL-DIC has been adopted for full-field deformation and strain measurements across multiple research areas, including studying metals and alloys deformation mechanisms and plasticity \cite{ni2025revisiting,ran2026comparative,ni2024automated,ni2025temperature,ozdur2021residual,gu2026electric,ni4836423automated,ni2025mapping,zhang5993534statistical,he2026mapping,hu2026oxygen,makinen2022detection}, laser welding of alloys \cite{pamarthi2023tailoring}, cyclic behaviors of shape memory alloys \cite{chen2023cyclic}, 
composites \cite{khouchani2024effect,mammadli2025universal,salian2021comparative,roach2025multiscale}, additive manufacturing and architected materials \cite{jirousek2023design,jirousek2024discovering,ye2024experimental,fry2023tensile,falta2018,li2025characterization}, soft materials and tissue mechanics \cite{summey2023open,roy2025toughening,sarkar2022quantification,pearce2024evaluation,tao2024experimental,zhou2025vaginal,yang2021smart,mcghee2023high,yang2026inertial,yang2025imac,bu2024high,wei2026machine}, civil/geophysics/wood and other construction materials \cite{harmal2023bioinspired,grubii2023impact,grubii2023measurement,grubii2023influence,falta2023mechanical,grubii2023quality,hafiz2025uniaxial}, batteries and energy materials \cite{wu2023deformation,courant2024design}, fluid, electromagnetic and other physics \cite{zhang2023pattern,zhang2025transitional,leon2022influence,zhang2025free}.  
It has also served as a benchmark for emerging DIC algorithms and has
been included in community evaluations \cite{YangR3DICnet,Ma21,FENG2024108267,kafka2024technique,YANG2025115908,gupta2024situ,feng2026stereo,Mangileva2024,PUREZA2026102532,schweickhardt2023digital,venter2025sun,wang2026interpretable,yang2022serialtrack,tong2026raftcorr} and reviews of open-source DIC
software \cite{jin2023recent,kopiika2024digital,naufal2024digital,kibrete2025free}. \\

However, the reference implementation of AL-DIC was developed in MATLAB, which limits its accessibility in environments without a MATLAB license and creates barriers to its integration into open-source Python-based scientific workflows. Commercial DIC codes, including  VIC-2D \cite{vic2d_software}, MatchID \cite{matchid_software}, GOM ARAMIS \cite{gom_aramis_software}, LaVision DaVis \cite{DaVis_software}, and EikoSim~\cite{EikoSim_software} provide graphical user interfaces and dedicated technical support. However, these packages are proprietary, and their internal implementations are generally not fully documented or publicly accessible.
Several open-source alternatives are also available. Ncorr
\cite{blaber2015ncorr} provides a MATLAB-based local subset implementation with a desktop graphical interface.
The Digital Image Correlation Engine (DICe) \cite{turner2015dice}
provides a C\texttt{++}-based correlation framework designed for large-scale computational workflows.
OpenCorr \cite{opencorr} is a C\texttt{++} library supporting accelerated two-dimensional, stereo, and volumetric correlation.
Python packages such as $\mu$DIC \cite{olufsen2020mudic}
and iCorrVision-2D \cite{icorrvision}, the latter of which includes an integrated graphical interface, provide open-source alternatives for DIC analysis.
However, these packages do not implement the complete AL-DIC workflow together with adaptive quadtree meshing and mask-aware subset treatment.
Table~\ref{tab:dic_codes} summarizes selected features of representative open-source DIC software packages. \\

This paper presents \textbf{pyALDIC}, an open-source Python implementation of AL-DIC that provides both an integrated graphical user interface and a scriptable Python application programming interface.
The software is distributed as a cross-platform, pip-installable package.
The AL-DIC formulation itself was introduced previously
\cite{yang2019augmented,yang2021fast};
the principal contributions of the present work are its accessible software implementation, integration of advanced analysis capabilities, and supporting software engineering.
Compared with the original MATLAB implementation of AL-DIC and other open-source DIC packages, pyALDIC provides the following features:

\begin{itemize}\setlength\itemsep{1pt}
\item[\textbf{(i)}] \textbf{Accessibility.} pyALDIC provides an open-source, cross-platform, GUI-enabled implementation of AL-DIC.
It is distributed through PyPI under the BSD-3-Clause license and includes user interfaces localized in eight languages.
\item[\textbf{(ii)}] \textbf{Adaptive and discontinuity-aware analysis.} The software combines adaptive quadtree meshing with five composable refinement criteria and automatic mask-aware subset splitting near cracks, holes, and ROI boundaries.
A single GUI control allows users to select either conventional local DIC or the complete AL-DIC formulation.
These capabilities improve the analysis of complex geometries, noisy images, and discontinuous displacement fields for which conventional untreated subsets may produce correlation artifacts.
\item[\textbf{(iii)}] \textbf{Computational performance and software engineering.} Numba just-in-time compilation enables the local solver to process approximately
$18{,}000$--$19{,}000$ points of interest per second in the representative benchmark tests reported in Table~\ref{tab:perf}.
The package also includes more than 1,300 unit and integration tests, automated continuous-integration testing on Linux for Python 3.10--3.12, reproducible examples, and comprehensive user documentation.

\end{itemize}

We will describe the above three advantages in Section~\ref{sec:software_des} in detail. The computing performance of pyALDIC is reported in Section~\ref{sec:perf}.  

\section{Software description}
\label{sec:software_des}

\subsection{Software architecture}
\label{sec:arch}

\begin{figure}[t!]
\centering
\includegraphics[width=1\linewidth]{ 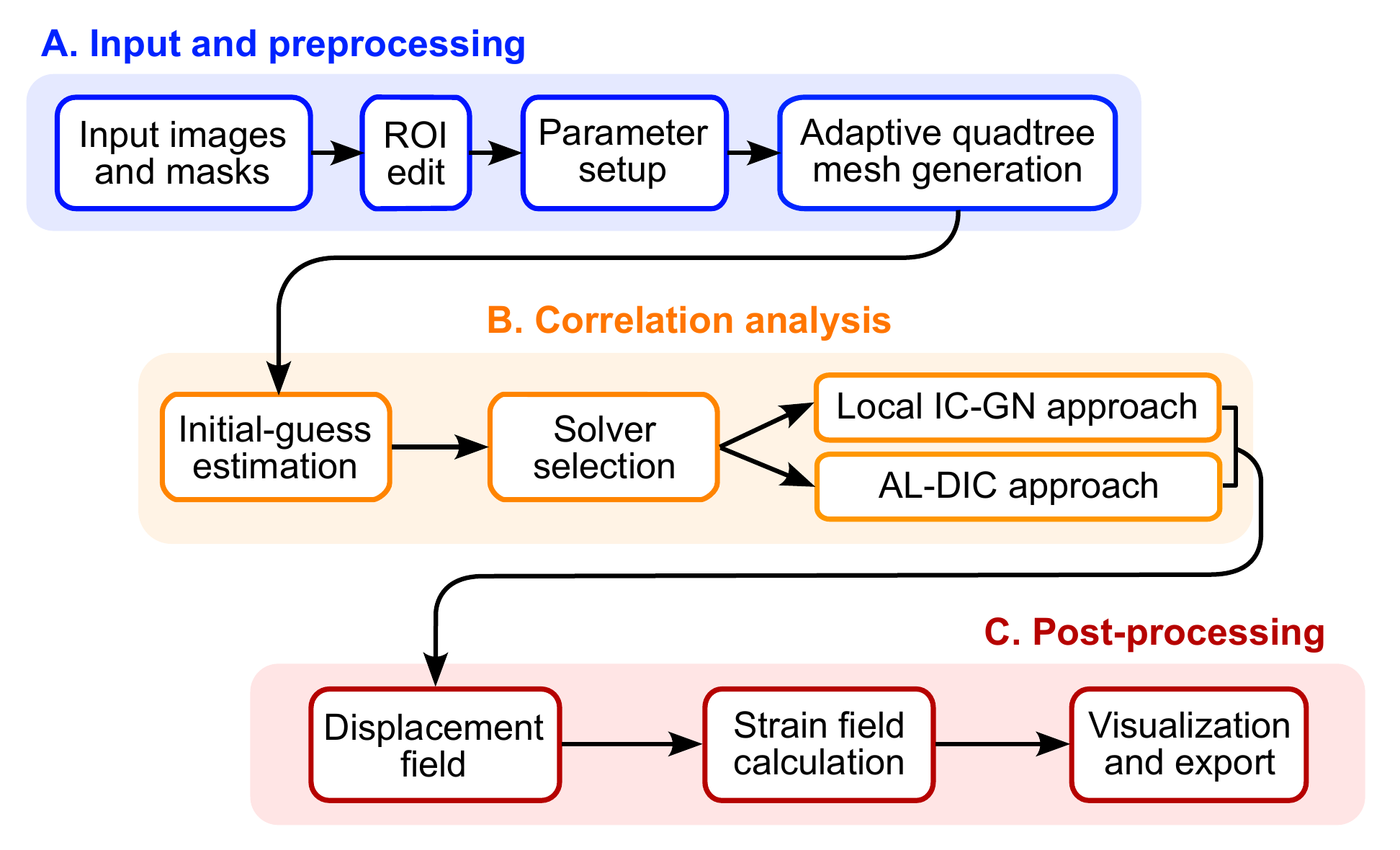}
\caption{pyALDIC processing workflow. An image sequence with optional masks passes through ROI editing, parameter configuration, adaptive quadtree meshing, and initial-guess estimation to a single-parameter \emph{solver choice} between \emph{Local IC-GN} and full \emph{AL-DIC} (IC-GN + Q4 FEM); both branches feed the strain computation and visualization. Every step is available through both the desktop GUI and the Python API.}
\label{fig:architecture}
\end{figure}

The pyALDIC processing workflow is summarized in Fig.~\ref{fig:architecture}. Users access the complete analysis pipeline through a PySide6 desktop application organized into three primary columns: an image browser, an interactive image canvas, and a combined parameter, execution, and results panel, as shown in Fig.~\ref{fig:gui}. Behind the graphical interface, an event-driven controller transfers user commands to the computational core, which contains dedicated modules for correlation, mesh generation, strain calculation, visualization, data export, and input/output operations. The computational core is decoupled from the graphical interface.
Consequently, all major operations available through the GUI can also be invoked directly through the Python API.
The complete workflow can therefore also be scripted for batch processing or executed within a Jupyter notebook without launching the desktop application.
The codebase comprises approximately 30 modules organized into domain-specific packages, including \texttt{core/}, \texttt{gui/}, \texttt{io/}, \texttt{mesh/}, \texttt{solver/}, \texttt{strain/}, \texttt{export/}, and \texttt{utils/}.  The software also includes more than 1,300 unit and integration tests implemented using \texttt{pytest}.\\

The processing pipeline reads DIC image sequences and their optional mask images in the formats of BMP, TIF, and PNG. It first constructs a Q4 finite-element mesh over the region of interest and applies adaptive quadtree refinement according to the selected refinement criteria.
An initial displacement estimate is then obtained using FFT-based cross-correlation \cite{landauer2018q}. The software subsequently executes either the Local IC-GN solver \cite{baker2004lucas,pan2013fast} or the complete AL-DIC formulation, in which the local correlation step is coupled to a global finite-element compatibility solve through ADMM iterations
\cite{yang2019augmented}.
For subsets that intersect cracks, holes, or ROI boundaries, pyALDIC applies automatic mask-aware subset splitting to exclude invalid pixels and prevent the correlation of independently moving material regions
\cite{poissant2010novel,yang2022staq,leu2026machine}. \\

The resulting displacement fields are spatially differentiated to obtain strain fields.
The GUI provides integrated tools for visualizing the calculated displacement and strain fields and exporting the results in multiple formats, including MATLAB \texttt{.mat}, NumPy \texttt{.npz}, CSV, PNG, GIF, MP4, and multipage PDF reports.

\subsection{Software functionalities}
\label{sec:func}

\subsubsection{Workflow and GUI}
\label{sec:gui}

\begin{figure}[h!]
\centering
\includegraphics[width=1\textwidth]{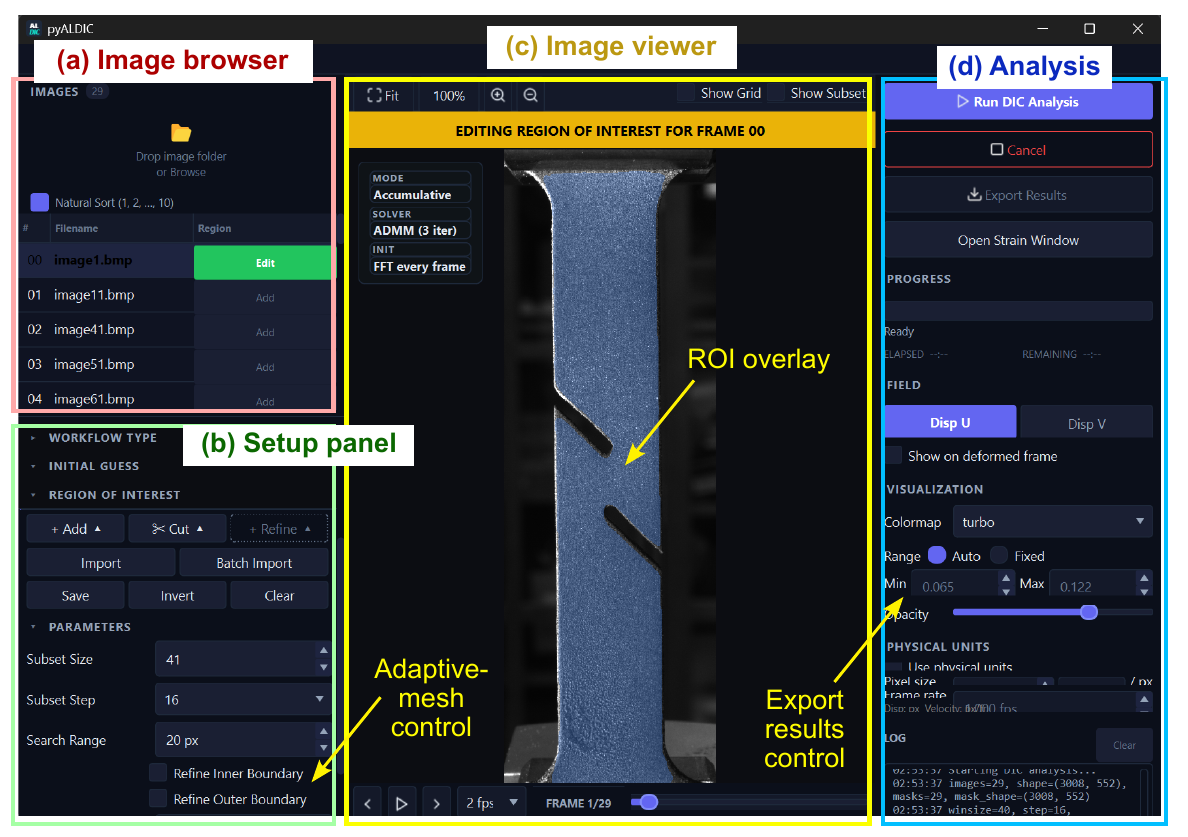}
\caption{Screenshot of the pyALDIC desktop GUI (v0.6.0) during ROI definition on the reference image of a uniaxial-tension experiment on a metallic specimen.
\textbf{(i)}~Left column: the uploaded image sequence, together with frame-specific ROI controls and collapsible panels for workflow type, initial-guess estimation, ROI editing, and DIC parameter configuration.
\textbf{(ii)}~Central canvas: the speckled specimen surface with the ROI overlay, analysis-configuration summary, mesh and subset overlays, zoom controls, and frame navigator.
\textbf{(iii)}~Right column: execution and export controls, progress and estimated-time-remaining indicators, displacement-field selection, visualization-range and colormap settings, physical-unit configuration, and the console log.}
\label{fig:gui}
\end{figure}

The pyALDIC GUI can be launched from a terminal using the command \texttt{al-dic}. As shown in Fig.~\ref{fig:gui}, the GUI uses a three-column layout that supports the analysis workflow without requiring user-written scripts. (i) The left column contains the image browser and allows users to select DIC images through file selection or drag-and-drop. It applies natural filename sorting and supports frame-specific ROI assignment for both accumulative and incremental tracking modes. 
(ii) The central \emph{canvas} provides image zooming and panning, an optional mesh overlay, subset visualization, and interactive drawing tools for adding, removing, and refining ROI regions through brushing or manual selection. 
(iii) The right column contains the DIC parameters and execution controls, including DIC window subset size, subset spacing step, initial-guess search range, mesh-refinement level, solver mode, run and cancellation controls,  progress and estimated-time-remaining indicators, visualization settings (e.g., colormap, opacity, deformed-configuration overlay), physical-unit conversion, and a console log. 
A demonstration of the complete workflow is distributed with the source repository and is also available online.\footnote{Demonstration video link: \url{https://www.youtube.com/watch?v=VEBfkVKMcKU}. Full walkthrough tutorials (English and Chinese) are available on YouTube (\url{https://www.youtube.com/watch?v=aLqDyM3tQII}; \url{https://www.youtube.com/watch?v=27TtJEMefNA}) and Bilibili (\url{https://www.bilibili.com/video/BV1dBjm6zEnQ}).} \\

All user-visible strings are wrapped in a Qt translation framework, and translation coverage is checked through continuous-integration tests.
Version~0.6.0 includes eight fully translated language locales:
English, Simplified Chinese, Traditional Chinese, Japanese, Korean, German, French, and Spanish.
Analysis results can be exported in MATLAB \texttt{.mat}, NumPy \texttt{.npz}, CSV, and PNG formats.
The software can also generate animated GIF files, MP4 videos, and multipage PDF reports containing field maps, line profiles, and analysis-parameter summaries.

\subsubsection{ALDIC formulation and post-processing analysis algorithms}
\label{sec:solver}

pyALDIC implements the AL-DIC mathematical formulation introduced in Ref.~\cite{yang2019augmented}, and extends the original implementation by incorporating adaptive quadtree meshing and mask-aware subset treatment for complex specimen geometries and discontinuous displacement fields \cite{yang2021fast,yang2022staq,leu2026machine}. Let the region of interest be denoted by $\Omega=\bigcup_i\Omega_i$, where $\Omega_i$ represents each local DIC subset using a piecewise-affine kinematic shape function. The coordinates of a material point in the deformed image, $\mathbf{y}$, are expressed in terms of the reference coordinates, $\mathbf{X}$, as
\begin{equation}
\mathbf{y}(\mathbf{X})=\mathbf{X}+\sum_i\big(\mathbf{u}_i+\mathbf{F}_i(\mathbf{X}-\mathbf{X}_{i0})\big)\,\chi_i(\mathbf{X}),
\label{eq:aldic_ansatz}
\end{equation}
where $\mathbf{X}_{i0}$ is the coordinates of the center of subset $\Omega_i$, $\mathbf{u}_i$ and $\mathbf{F}_i$ are its averaged displacement and  displacement gradient tensor at $\mathbf{X}_{i0}$.  $\chi_i$ is the indicator function of $\Omega_i$.  $\chi_i=1$ if a point is within subset $\Omega_i$; otherwise $\chi_i=0$ if not. Each subset has independent local variables $\{\mathbf{u}_i,\mathbf{F}_i\}$. Thus, the locally resolved displacement and displacement-gradient fields are not generally kinematically compatible across neighboring subsets. AL-DIC restores compatibility by introducing an auxiliary, globally continuous displacement field $\hat{\mathbf{u}}$, represented on a finite-element mesh.  For adaptive quadtree meshes, the finite-element representation includes constraints for hanging nodes at transitions between different refinement levels.
The local and global variables are coupled through the kinematic constraints

\begin{equation}
\mathbf{F}_i=\nabla\hat{\mathbf{u}}(\mathbf{X}_{i0}),\qquad \mathbf{u}_i=\hat{\mathbf{u}}(\mathbf{X}_{i0}).
\label{eq:aldic_constraint}
\end{equation}

The DIC correlation objective minimizes the sum of squared grayscale-intensity differences subject to the kinematic constraints in Eqs.~\eqref{eq:aldic_ansatz} and \eqref{eq:aldic_constraint}.
Using an augmented-Lagrangian formulation, the constrained objective is written as
\begin{equation}
\begin{split}
\mathcal{L}_0=\sum_i\int_{\Omega_i}\Big(
&\big|f(\mathbf{X})-g\big(\mathbf{X}+\mathbf{u}_i+\mathbf{F}_i(\mathbf{X}-\mathbf{X}_{i0})\big)\big|^2 \\[2pt]
&+\tfrac{\beta}{2}\big|(\mathbf{D}\hat{\mathbf{u}})_i-\mathbf{F}_i\big|^2
+\boldsymbol{\nu}_i:\big((\mathbf{D}\hat{\mathbf{u}})_i-\mathbf{F}_i\big)\\[2pt]
&+\tfrac{\mu}{2}\big|\hat{\mathbf{u}}_i-\mathbf{u}_i\big|^2
+\boldsymbol{\lambda}_i\cdot\big(\hat{\mathbf{u}}_i-\mathbf{u}_i\big)\Big)\,\mathrm{d}\mathbf{X},
\end{split}
\label{eq:aldic_main}
\end{equation}
where $f$ and $g$ denote the grayscale intensities of the reference and deformed images, respectively. The operator $\mathbf{D}$ denotes the discrete spatial-gradient operator applied to $\hat{\mathbf{u}}$;
$\boldsymbol{\nu}_i$ and $\boldsymbol{\lambda}_i$ are Lagrange multipliers associated with the gradient and displacement constraints, respectively; and
$\beta,\mu>0$ are penalty parameters. The symbol $|\cdot|$ denotes the Frobenius norm, $(:)$ the double inner product between tensors, and $(\cdot)$ the inner product between two vectors. \\

Equation~\eqref{eq:aldic_main} is minimized using the alternating direction method of multipliers (ADMM) \cite{boyd2011distributed} that alternates a local, per-subset IC-GN update of $\{\mathbf{u}_i\}$, a global finite difference or finite element-based solve for $\hat{\mathbf{u}}$, and a multiplier update, as summarized in Algorithm~\ref{alg:aldic}. 
Following the original AL-DIC derivation \cite{yang2019augmented}, the scaled dual variables are defined as $\mathbf{W}_i=\boldsymbol{\nu}_i/\beta$
and $\mathbf{v}_i=\boldsymbol{\lambda}_i/\mu$.
The local subproblem is accelerated by setting $\mathbf{F}_i^{k+1}=(\mathbf{D}\hat{\mathbf{u}}^k)_i$, such that only the subset displacement variables $\{\mathbf{u}_i\}$ are updated through IC-GN.
The complete derivation of the AL-DIC formulation is provided in Ref.~\cite{yang2019augmented} and the flowchart is summarized in Algorithm~\ref{alg:aldic}. \\

pyALDIC extends the global compatibility solve from a uniform Q4 mesh to an adaptively refined quadtree mesh.
Hanging-node constraints are introduced for the midside degrees of freedom that occur at transitions between refinement levels
\cite{yang2021fast,yang2022staq}.
The GUI reports the conventional odd-valued full subset width,
$2h+1$ pixels.
For example, a half-width of $h=20$ pixels corresponds to a displayed subset size of 41~pixels.
Internally, the Python API uses the even-valued parameter
\texttt{winsize}$=2h$,
which equals 40~pixels in this example.
Thus, the GUI and API representations differ by one pixel but describe the same correlation window convention.

\begin{algorithm}[t!]
\small
\SetAlgoLined
\KwIn{Reference image $f$, deformed image $g$, ROI mesh $\mathcal{M}$, mask, penalty parameters $\beta,\mu$, convergence tolerance $\varepsilon$}
\textbf{Step 1.} Construct the DIC mesh $\mathcal{M}$ and apply adaptive quadtree refinement\;
\textbf{Step 2.} Compute the initial displacement estimate
$\hat{\mathbf{u}}^0$
using FFT-based cross-correlation with an automatically expanded search range\;
\textbf{Step 3.} Initialize the scaled dual variables
$\mathbf{W}^0=\mathbf{0}$ and $\mathbf{v}^0=\mathbf{0}$,
and set iteration number $k=0$\;
\While{$\|\hat{\mathbf{u}}^{k+1}-\hat{\mathbf{u}}^{k}\| > \varepsilon$ \textbf{and} $k < k_{\max}$}{
  \emph{\textbf{Subproblem~1 (Local IC-GN update)}}: set $\mathbf{F}^{k+1}=\mathbf{D}\hat{\mathbf{u}}^{k}$\;
  \For{each DIC mesh node $i \in \mathcal{M}$}{
    Intersect subset $i$ with the ROI mask and retain the connected valid component containing the subset center\;
    Apply mask-aware subset splitting when the subset intersects a displacement discontinuity or an ROI boundary\;
    Compute
    $\mathbf{u}_i^{k+1} \leftarrow \argmin_{\mathbf{u}_i}\big\{ \mathrm{SSD}_i(\mathbf{u}_i) + \tfrac{\mu}{2}\|\hat{\mathbf{u}}_i^{k} - \mathbf{u}_i + \mathbf{v}_i^{k}\|^2 \big\}$ using IC-GN, where $\mathrm{SSD}_i(\mathbf{u}_i)$ is the first term on the right side of Eq~(\ref{eq:aldic_main})\;
  }
  \emph{\textbf{Subproblem~2 (Global compatibility solve)}}\;
  Solve global displacement variable $\hat{\mathbf{u}}^{k+1}$: $\big( \beta\,\mathbf{D}^{\!\top}\mathbf{D} + \mu\,\mathbf{I} \big)\hat{\mathbf{u}}^{k+1} = \beta\,\mathbf{D}^{\!\top}\!\big(\mathbf{F}^{k+1}-\mathbf{W}^{k}\big) + \mu\big(\mathbf{u}^{k+1}-\mathbf{v}^{k}\big)$ on $\mathcal{M}$\;
  \emph{\textbf{Dual variables update}}\;
  $\mathbf{W}^{k+1} \leftarrow \mathbf{W}^{k} + \big(\mathbf{D}\hat{\mathbf{u}}^{k+1} - \mathbf{F}^{k+1}\big)$,\quad $\mathbf{v}^{k+1} \leftarrow \mathbf{v}^{k} + \big(\hat{\mathbf{u}}^{k+1} - \mathbf{u}^{k+1}\big)$\;
  Update iteration number $k\leftarrow k+1$\;
}
\KwOut{Local subset displacement vectors $\mathbf{u}$ and globally compatible displacement vectors $\hat{\mathbf{u}}$
}
\caption{ AL-DIC implemented in pyALDIC. 
Disabling the global compatibility step reduces the procedure to conventional Local DIC.}
\label{alg:aldic}
\end{algorithm}

\begin{figure}[h!]
\centering
\includegraphics[width=\textwidth]{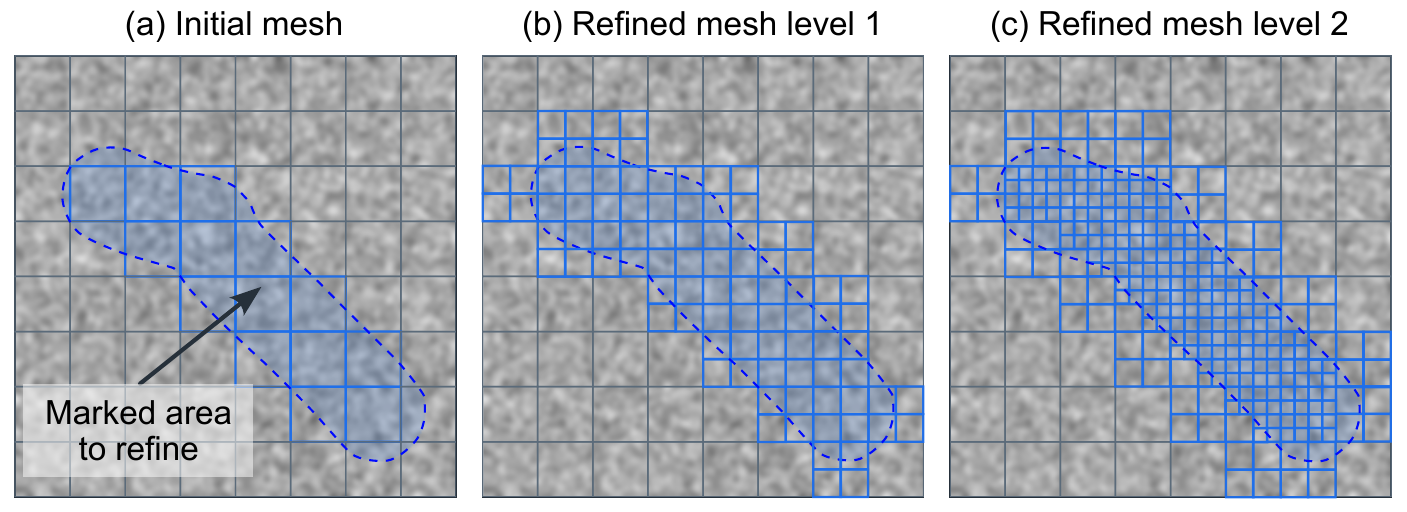}
\caption{Adaptive quadtree refinement through two successive refinement levels.
\textbf{(a)}~Uniform base mesh with the user-selected refinement region indicated.
\textbf{(b)}~Mesh after one level of quadtree subdivision.
\textbf{(c)}~Mesh after two levels of subdivision.
Flagged elements are recursively divided into four child elements, thereby concentrating correlation nodes in the selected region while retaining a coarser discretization elsewhere.
}
\label{fig:adaptive_mesh}
\end{figure}

\paragraph{Adaptive quadtree mesh refinement} 
Adaptive quadtree refinement for DIC analysis near complex geometries and displacement discontinuities was introduced in the spatiotemporally adaptive quadtree (STAQ) framework~\cite{yang2022staq}. pyALDIC provides an interactive, GUI-enabled implementation of this refinement strategy. Starting from a uniform Q4 base mesh, selected elements are recursively subdivided into four child elements, as illustrated in Fig.~\ref{fig:adaptive_mesh}.
Refinement can be triggered by one or more of five criteria:
(i) \emph{mask boundary},
(ii) \emph{ROI boundary},
(iii) \emph{brush-defined region},
(iv) \emph{manual selection},
and
(v) \emph{a posteriori correlation error}.
For the error-based criterion, elements are refined when the IC-GN residual exceeds a prescribed threshold after an initial outer ADMM iteration \cite{yang2021fast}.
The refinement criteria can be combined, and the GUI provides
\texttt{Light},
\texttt{Medium}, and
\texttt{Strong}
presets that apply predefined combinations of these criteria.

\begin{figure}[h!]
\centering
\includegraphics[width=\textwidth]{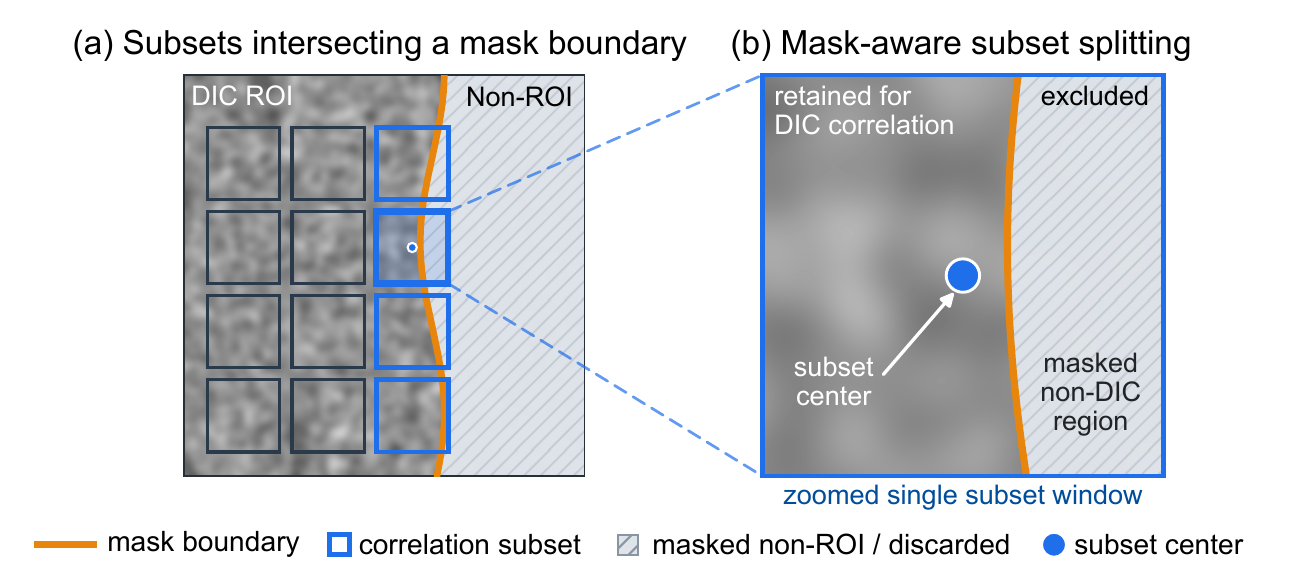}
\caption{Mask-aware subset splitting near a displacement discontinuity.
(a) Correlation subsets near a mask boundary overlap both valid speckle regions and the masked region.
(b) The highlighted subset is intersected with the mask, and only the connected valid component containing the subset center is retained for the IC-GN calculation.
Masked pixels and disconnected valid regions on the opposite side of the discontinuity are excluded.
}
\label{fig:window_splitting}
\end{figure}

\paragraph{Mask-aware subset splitting} 
A conventional square correlation subset that intersects a mask boundary, such as a crack face or the boundary of a hole, may contain both invalid pixels and valid speckle patterns belonging to independently moving material regions.
Directly correlating such a subset may produce an erroneous displacement estimate because no single affine transformation can represent the motion of all included pixels.
For each correlation node, pyALDIC intersects the subset with the ROI mask and identifies the connected valid components using connected-component labeling.
Only the component containing the subset center is retained, and the local IC-GN problem is solved using this valid pixel set, as illustrated in Fig.~\ref{fig:window_splitting}.
Consequently, a subset adjacent to a crack correlates only the speckle pattern located on the same crack face as the subset center and does not mix information from the opposite face.
If the retained component contains fewer than one-half of the pixels in the original subset, the corresponding correlation node is flagged as unreliable.
This procedure improves displacement estimation in the region where conventional subsets would otherwise straddle a discontinuity.
A worked Mode-I crack example demonstrating this capability is distributed with the software.

\begin{figure}[h!]
\centering
\includegraphics[width=\textwidth]{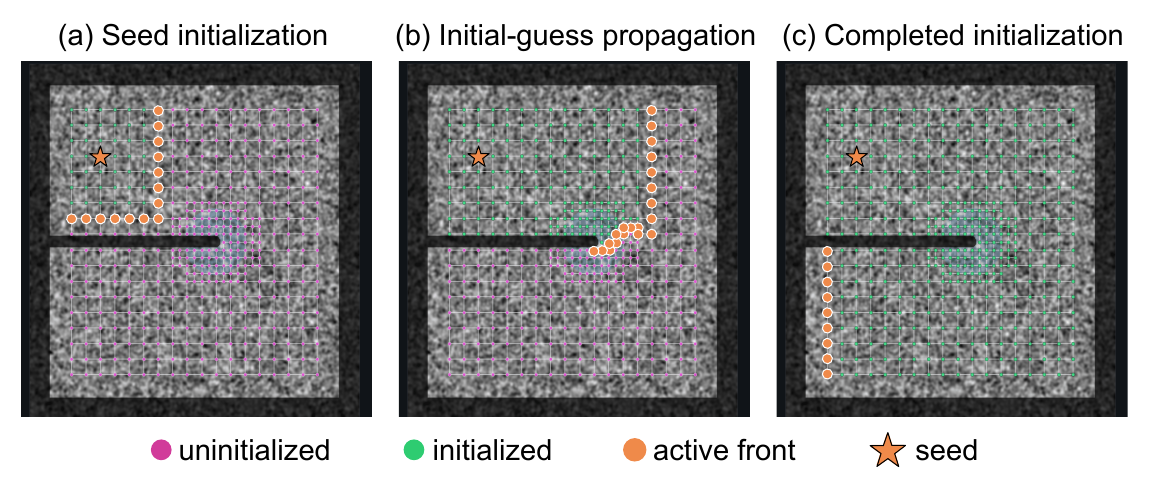}
\caption{Seed-propagation initial-displacement estimation for a cracked specimen.
 {(a)}~Initialization from a user-defined seed point.
 {(b)}~Intermediate propagation stage.
 {(c)}~Completed initial-displacement field.
Solved nodes are shown in green, the active propagation front in orange, unsolved nodes in magenta, and the initial seed by a star.
Propagation follows mask-connected paths and therefore respects the specimen geometry and displacement discontinuities.
}
\label{fig:seed_propagation}
\end{figure}

\paragraph{Seed-propagation initial guess} 
As an alternative to whole-field FFT-based cross-correlation \cite{landauer2018q}, pyALDIC provides a seed-propagation strategy for estimating initial displacements in cases involving large inter-frame motion, complex geometries, or discontinuous displacement fields
(Fig.~\ref{fig:seed_propagation}).
The user specifies one or more seed points, with at least one seed assigned to each disconnected mask region.
The displacement at each seed is initialized through a single-point cross-correlation calculation.
The estimate is subsequently propagated to neighboring mesh nodes using first-order extrapolation informed by the local displacement gradient
$\mathbf{F}$.
Propagation is restricted to mask-connected paths, allowing the initial-displacement field to follow specimen boundaries and propagate around cracks and holes without crossing discontinuities.

\paragraph{Selection between Local DIC and AL-DIC} 
A single GUI control switches between \emph{Local DIC} and \emph{AL-DIC}.
Local DIC performs only the local IC-GN update represented by Subproblem~1 in Algorithm~\ref{alg:aldic}, without applying the global compatibility solve.
It is therefore computationally less expensive and is well suited to smooth, low-noise displacement fields with modest displacement gradients. 
AL-DIC executes the complete local--global iteration in Algorithm~\ref{alg:aldic}.
The global finite-element or finite-difference solve imposes compatibility and regularizes the displacement field.
As established for the original AL-DIC formulation
\cite{yang2019combining,yang2019augmented},
this coupling becomes particularly advantageous as image noise increases and when displacement fields contain strong gradients or discontinuities.

\subsubsection{Performance and distribution}
\label{sec:perf}

Computationally intensive kernels, including IC-GN Hessian matrix evaluation, chunked normalized cross-correlation searches, and mesh interpolation,  are just-in-time (JIT) compiled by Numba \cite{lam2015numba}. The first execution includes compilation overhead, whereas subsequent executions use the compiled kernels.

Representative performance was measured after JIT warm-up on a 24-thread Intel Raptor Lake CPU using Python~3.13, NumPy~2.1, and Numba~0.61. The benchmark used a subset size of 31 pixels $\times$ 31 pixels, a subset spacing of 8~pixels,
and an FFT-based initial-displacement estimate. For image sizes ranging from
512 pixels $\times$ 512 pixels to 1024 pixels $\times$ 1024 pixels, Local DIC processed approximately $18{,}000$--$19{,}000$ points of interest per second.

The complete AL-DIC solver was slower compared to Local DIC because each ADMM iteration included an additional global finite-element compatibility solve, and it may take 3--5 ADMM iteration steps to converge. Under the representative convergence settings used for these tests, the measured throughput difference was approximately $2\times$ for the $256\times256$-pixel example and increased to approximately $6\times$ for the $1024\times1024$-pixel example (Table~\ref{tab:perf}). For the smallest image, the relative cost of fixed finite-element assembly and solver initialization was more pronounced.

The chunked normalized cross-correlation implementation introduced in version~0.4 also reduced the peak memory required for the FFT-based initial guess. For a
$4096\times4096$-pixel benchmark, peak memory usage decreased from approximately 37~GB to 12~GB, corresponding to a reduction of approximately threefold.
This modification eliminated an out-of-memory failure that previously occurred for large images and search ranges. The benchmark scripts used to reproduce the throughput and memory measurements are provided in the repository's \texttt{tools/} directory. \\

\begin{table}[t!]
\small
\centering
\caption{Representative computational throughput measured after JIT warm-up on a 24-thread Intel Raptor Lake CPU. The benchmark used a $31\times31$-pixel subset, an 8-pixel subset spacing, and an FFT-based initial-displacement estimate.
Throughput is reported in points of interest per second (POI\,$\cdot$\,s$^{-1}$).
.}
\label{tab:perf}
\begin{tabular}{cccc}
\toprule
Image size (pixels) & \# of points & Local DIC (POI\,$\cdot$\,s$^{-1}$) & AL-DIC (POI\,$\cdot$\,s$^{-1}$) \\
\midrule
$256 \times 256$  & 484    & 6{,}000  & 2{,}800 \\
$512 \times 512$  & 2{,}916  & 18{,}300 & 3{,}900 \\
$1024 \times 1024$ & 13{,}924 & 18{,}600 & 2{,}900 \\
\bottomrule
\end{tabular}
\end{table}

pyALDIC is released under the BSD-3-Clause license through the following channels: (i) PyPI (pyALDIC can be installed from PyPI using ``\texttt{pip install al-dic}''), (ii) GitHub Release wheels support offline environments. For development, the source package can be installed in editable mode using ``\texttt{pip install -e ".[dev]"}''. The continuous-integration workflow, implemented using GitHub Actions, executes the complete test suite on Linux with Python versions 3.10--3.12.
The package supports Windows, macOS, and Linux and uses the same installation and analysis workflow across these operating systems.

\section{Conclusions}
\label{sec:impact}

We have presented pyALDIC, an open-source, cross-platform Python implementation of AL-DIC~\cite{yang2019augmented} that provides both a graphical user interface and a scriptable Python API.
The software integrates adaptive quadtree meshing, automatic mask-aware subset splitting for complex geometries \cite{yang2022staq,leu2026machine} and discontinuous displacement fields \cite{poissant2010novel,yang2022staq}, and selectable Local DIC and AL-DIC solver modes.  The AL-DIC formulation's capability was evaluated and compared with other popular DIC codes using the DIC Challenge~2.0 datasets~\cite{dicchallenge2exme}. The capabilities of AL-DIC were also demonstrated using synthetic benchmarks \cite{yang2019augmented,yang2021fast,yang2022staq,leu2026machine} and experimental examples as reported in the literature \cite{ni2025revisiting,ran2026comparative,yang2019augmented,yang2021fast,yang2022staq,leu2026machine,ni2024automated,ni2025temperature,ozdur2021residual,gu2026electric,ni4836423automated,ni2025mapping,zhang5993534statistical,he2026mapping,hu2026oxygen,makinen2022detection,pamarthi2023tailoring,chen2023cyclic,khouchani2024effect,mammadli2025universal,salian2021comparative,roach2025multiscale,jirousek2023design,jirousek2024discovering,ye2024experimental,fry2023tensile,falta2018,li2025characterization,summey2023open,roy2025toughening,sarkar2022quantification,pearce2024evaluation,tao2024experimental,zhou2025vaginal,yang2021smart,mcghee2023high,yang2026inertial,yang2025imac,bu2024high,wei2026machine,harmal2023bioinspired,grubii2023impact,grubii2023measurement,grubii2023influence,falta2023mechanical,grubii2023quality,hafiz2025uniaxial,wu2023deformation,courant2024design,zhang2023pattern,zhang2025transitional,leon2022influence,zhang2025free,tong2026raftcorr}.

The parameter definitions, ROI conventions, refinement settings, and error-reporting outputs in pyALDIC are designed to be consistent with the iDICs \emph{Good Practices Guide for Digital Image Correlation}~\cite{idics_gpg_2018,idics_gpg_2025}.
This alignment reduces the learning curve for practitioners familiar with established DIC terminology and analysis workflows.



Future development will continue targeting (i) extending pyALDIC to three-dimensional stereo-DIC based on the recently developed stereo AL-DIC formulation~\cite{tong2025stereo3d}, (ii) accelerating the IC-GN computational kernels using GPU computing, including Numba CUDA, and (iii) improving integration with computational mechanics workflows for experimental--computational field comparison, constitutive-model calibration, and data-driven model discovery \cite{wei2026machine,jin2023recent,flaschel2021unsupervised,you2022physics,pearce2022combining,kirchhoff2024inference,wihardja2025constitutive,kumar2025comparative,spencer2026full}. 

The source code, datasets, and documentation of pyALDIC are openly available at \url{https://github.com/zachtong/pyALDIC}. Contributions, feature requests, and issue reports are welcome through the project's GitHub Issues and Discussions pages.

\section*{CRediT authorship contribution statement}
\textbf{Zixiang Tong:} Conceptualization, Methodology, Software, Validation, Formal analysis, Investigation, Data curation, Visualization, Writing -- original draft, Writing -- review \& editing. \textbf{Jin Yang:} Conceptualization, Methodology, Visualization, Supervision, Resources, Funding acquisition, Project administration, Writing -- review \& editing.

\section*{Conflict of interest}
The authors declare no conflict of interest. There has been no significant financial support for this work that could have influenced its outcome.

\section*{Data availability}
Runnable examples, together with the corresponding datasets and analysis scripts, are distributed with the source repository and archived with the software on Zenodo~\cite{pyaldic_zenodo}.
These materials include a synthetic accuracy benchmark, a Mode~I crack example analyzed using mask-aware subset splitting, an adaptive-mesh refinement example, and an experimental uniaxial-tension dataset.
The experimental images were acquired by the authors and are released under the same license as the software.

\section*{Declaration of generative AI and AI-assisted technologies in the manuscript preparation process}
During the preparation of this work, the authors used Anthropic Claude and OpenAI ChatGPT to polish the manuscript text and support the preparation of schematic figures through author-reviewed Python code. The authors reviewed and edited all AI-assisted content and take full responsibility for the content of the publication.

\section*{Acknowledgements}
Z.T. acknowledges support from the University Graduate Continuing Fellowship, the Graduate Excellence Fellowship, and the Professional Development Award from the Cockrell School of Engineering at The University of Texas at Austin.
J.Y. gratefully acknowledges support from the U.S. National Science Foundation under Grants No.~2232428, 2441460, and 2452029; the U.S. Office of Naval Research under Grant No.~N000142612190 through Dr. Timothy Bentley; and a Pilot Seed Gift from the Semiconductor Research Corporation.

\bibliographystyle{unsrt}
\bibliography{reference}

@article{ni2024automated,
  title={Automated analysis framework of strain partitioning and deformation mechanisms via multimodal fusion and computer vision},
  author={Ni, Ran and Boehlert, Carl J and Zeng, Ying and Chen, Bo and Huang, Saijun and Zheng, Jiang and Zhou, Hao and Wang, Qudong and Yin, Dongdi},
  journal={International Journal of Plasticity},
  volume={182},
  pages={104119},
  year={2024},
  publisher={Elsevier}
}

@article{pamarthi2023tailoring,
  title={Tailoring the weld microstructure to prevent solidification cracking in remote laser welding of AA6005 aluminium alloys using adjustable ringmode beam},
  author={Pamarthi, Venkat Vivek and Sun, Tianzhu and Das, Abhishek and Franciosa, Pasquale},
  journal={journal of Materials Research and Technology},
  volume={25},
  pages={7154--7168},
  year={2023},
  publisher={Elsevier}
}

@article{ni2025temperature,
  title={Temperature-dependent interplay of intra-and inter-granular deformation mechanisms in {Mg-10Y: Statistical analysis from an HRDIC perspective}},
  author={Ni, Ran and Boehlert, Carl J and Chen, Bo and Zhang, Yuanshuai and Zeng, Ying and Zheng, Jiang and Zhou, Hao and Wang, Qudong and Yin, Dongdi},
  journal={Acta Materialia},
  volume={296},
  pages={121256},
  year={2025},
  publisher={Elsevier}
}

@article{ni2025revisiting,
  title={Revisiting tension-compression asymmetry in a {M}g alloy: insights from statistical strain partitioning and intra-/inter-granular mechanisms at the nanoscale},
  author={Ni, Ran and Boehlert, Carl J and Zheng, Xianhua and Ran, Yaming and Huang, Saijun and Zeng, Ying and Zheng, Jiang and Wang, Qudong and Zhou, Hao and Yin, Dongdi},
  journal={International Journal of Plasticity},
  pages={104463},
  year={2025},
  publisher={Elsevier}
}

@article{chen2023cyclic,
  title={Cyclic behaviours of superelastic shape-memory alloy plates joined by tungsten inert gas welding},
  author={Chen, Zhi-Peng and Zhu, Songye},
  journal={Construction and Building Materials},
  volume={402},
  pages={132768},
  year={2023},
  publisher={Elsevier}
}

@article{ozdur2021residual,
  title={Residual intensity as a morphological identifier of twinning fields in microscopic image correlation},
  author={{\"O}zd{\"u}r, NA and {\"U}{\c{c}}el, IB and Yang, J and Ayd{\i}ner, CC},
  journal={Experimental Mechanics},
  volume={61},
  number={3},
  pages={499--514},
  year={2021},
  publisher={Springer}
}

@article{gu2026electric,
  title={Electric current-driven heterogeneous microstructures in dual-phase titanium alloys},
  author={Gu, Shaojie and Kimura, Yasuhiro and Cui, Yi and Morita, Yasuyuki and Isoi, Sora and Liu, Chang and Yan, Xinming and Ju, Bingfeng and Yang, Huayong and Toku, Yuhki and others},
  journal={Nature Communications},
  volume={17},
  number={1},
  pages={3470},
  year={2026},
  publisher={Nature Publishing Group UK London}
}

@article{ni4836423automated,
  title={Automated Analysis Framework of Strain Partitioning and Deformation Mechanisms Via Hrdic-Ebsd Fusion and Computer Vision: Application to a {M}g Alloy},
  author={Ni, Ran and Boehlert, Carl J and Zeng, Ying and Chen, Bo and Huang, Saijun and Zheng, J and Zhou, Hao and Wang, Qudong and Yin, Dongdi},
  journal={Available at SSRN 4836423}
}

@article{ni2025mapping,
  title={Mapping the strain-localization evolution of grain boundary and its interactions with slip/twin at the microscale},
  author={Ni, Ran and Huang, Saijun and Fan, Lingling and Wei, Kang and Zeng, Ying and Zheng, Jiang and Wang, Qudong and Zhou, Hao and Yin, Dongdi},
  journal={Journal of Magnesium and Alloys},
  year={2025},
  publisher={Elsevier}
}

@article{zhang5993534statistical,
  title={Statistical assessment of slip transfer in a {M}g alloy: insights from nano-scale deformation-field continuity via {HRDIC}},
  author={Zhang, Yuanshuai and Boehlert, Carl J and Ni, Ran and Chen, Bo and Zeng, Ying and Zheng, J and Wang, Qudong and Zhou, Hao and Yin, Dongdi},
  journal={Available at SSRN 5993534}
}

@article{he2026mapping,
  title={{Mapping precipitate-slip/twin interactions in Mg-Sn alloys: Insights from strain partitioning via HRDIC-EBSD multimodal analysis}},
  author={He, Siyu and Zeng, Ying and Huang, Meilin and Ni, Ran and Zhang, Yuanshuai and Jiang, Bin and Yin, Dongdi},
  journal={Scripta Materialia},
  volume={277},
  pages={117227},
  year={2026},
  publisher={Elsevier}
}

@article{ran2026comparative,
  title={{A comparative study of tension and compression in pure Mg sheet: Unveiling nanoscale intra-and inter-granular accommodation mechanisms via HRDIC}},
  author={Ran, Yaming and Zhang, Yuanshuai and Hua, Shen and Zheng, Jiang and Wang, Qudong and Zhou, Hao and Ni, Ran and Yin, Dongdi},
  journal={Journal of Materials Science \& Technology},
  year={2026},
  publisher={Elsevier}
}

@article{hu2026oxygen,
  title={{Oxygen effect on deformation mechanism and mechanical behavior of pure Ti at 77K}},
  author={Hu, Jiajun and Zhang, Dongmei and Zhao, Yonghao and Caron, Arnaud and Zhang, Yuanshuai and Wang, Shuaizhuo and Gao, Bo and Xiao, Lirong and Yin, Dongdi and Zhou, Hao and others},
  journal={Acta Materialia},
  pages={122433},
  year={2026},
  publisher={Elsevier}
}

@article{jirousek2023design,
  title={Design exploration of additively manufactured chiral auxetic structure using explainable machine learning},
  author={Jirousek, Ondrej and Palar, Pramudita Satria and Falta, Jan and Dwianto, Yohanes Bimo and others},
  journal={Materials \& Design},
  volume={232},
  pages={112128},
  year={2023},
  publisher={Elsevier}
}

@article{khouchani2024effect,
  title={{Effect of cellulose nanocrystals on performance of PVA fiber-reinforced geopolymer composites: Reaction kinetics, bending behavior, and toughening mechanisms}},
  author={Khouchani, Oussama and Harmal, Anass and El-Korchi, Tahar and Tao, Mingjiang and Walker, Harold W},
  journal={Construction and Building Materials},
  volume={435},
  pages={136727},
  year={2024},
  publisher={Elsevier}
}

@article{jirousek2024discovering,
  title={{Discovering chiral auxetic structures with near-zero Poisson's ratio using an active learning strategy}},
  author={Jirousek, Ondrej and Falta, Jan and Dwianto, Yohanes Bimo and Palar, Pramudita Satria and others},
  journal={Materials \& Design},
  volume={244},
  pages={113133},
  year={2024},
  publisher={Elsevier}
}

@article{makinen2022detection,
  title={Detection of the onset of yielding and creep failure from digital image correlation},
  author={M{\"a}kinen, Tero and Zaborowska, Agata and Frelek-Kozak, Ma{\l}gorzata and J{\'o}{\'z}wik, Iwona and Kurpaska, {\L}ukasz and Papanikolaou, Stefanos and Alava, Mikko J},
  journal={Physical Review Materials},
  volume={6},
  number={10},
  pages={103601},
  year={2022},
  publisher={APS}
}

@article{mammadli2025universal,
  title={Universal characteristics of local strain fields for creep failure prediction},
  author={Mammadli, Bakhtiyar and M{\"a}kinen, Tero and Frydrych, Karol and Asteris, Panagiotis G and Papanikolaou, Stefanos},
  journal={International Journal of Mechanical Sciences},
  pages={110612},
  year={2025},
  publisher={Elsevier}
}

@article{ye2024experimental,
  title={{Experimental Evaluation of the Effects of Discrete-Grading-Induced Discontinuities on the Material Properties of Functionally Graded Ti-6Al-4V Lattices}},
  author={Ye, Junyang and Babazadeh-Naseri, Ata and Higgs III, C Fred and Fregly, Benjamin J},
  journal={Materials},
  volume={17},
  number={4},
  pages={822},
  year={2024},
  publisher={MDPI}
}

@article{fry2023tensile,
  title={Tensile property measurement of lattice structures},
  author={Fry, AT and Crocker, LE and Lodeiro, MJ and Poole, M and Woolliams, P and Koko, A and Leung, N and England, D and Breheny, C},
  year={2023}
}

@incollection{salian2021comparative,
  title={Comparative Study of Strain Using 2D {Digital Image Correlation} and Extensometer on Glass Fiber},
  author={Salian, Abhilash and Ashwini, TP},
  booktitle={Smart Sensors Measurements and Instrumentation: Select Proceedings of CISCON 2020},
  pages={351--366},
  year={2021},
  publisher={Springer}
}

@inproceedings{falta2018,
  title={{Strain-rate dependent compressive properties of inverted honeycomb lattice and bulk cylindrical samples 3D printed by MSLA method}},
  author={Falta, Jan and Yulmaz, Yunus Emre},
  booktitle={27/28th Internatinoal Conference Engineering Mechsnics},
  pages={93--96},
  year={2022},
  organization={},
  doi={10.21495/51293},
}

@inproceedings{roach2025multiscale,
  title={A Multiscale Modeling Approach for Progressive Damage Analysis of Notched Ceramic Matrix Composites},
  author={Roach, James and Zhang, Dianyun},
  booktitle={AIAA SCITECH 2025 Forum},
  pages={1361},
  year={2025}
}

@inproceedings{li2025characterization,
  title={Characterization of Stress Localization in Epoxy Materials Using Augmented {L}agrangian Digital Image Correlation},
  author={Li, Ming and Al Bataineh, Ali},
  booktitle={International Conference on WorldS4},
  pages={191--198},
  year={2025},
  organization={Springer}
}

@article{summey2023open,
  title={Open source, In-Situ, intermediate Strain-Rate tensile impact device for soft materials and cell culture systems},
  author={Summey, Luke and Zhang, Jing and Landauer, AK and Sergay, Jamie and Yang, Jin and Daul, Annalise and Tao, Jialiang and Park, Jessica and McGhee, A and Franck, C},
  journal={Experimental Mechanics},
  volume={63},
  number={9},
  pages={1445--1460},
  year={2023},
  publisher={Springer}
}

@article{roy2025toughening,
  title={Toughening Starch-Based Bioplastics with Soy Amyloid Fibrils Produced from Tofu Wastewater},
  author={Roy Goswami, Shrestha and Mykolenko, Svitlana and Kong, Xiang and Mezzenga, Raffaele},
  journal={ACS Sustainable Chemistry \& Engineering},
  volume={13},
  number={21},
  pages={8068--8077},
  year={2025},
  publisher={ACS Publications}
}

@article{sarkar2022quantification,
  title={Quantification of errors in applying {DIC} to fiber networks imaged by confocal microscopy},
  author={Sarkar, M and Notbohm, J},
  journal={Experimental Mechanics},
  volume={62},
  number={7},
  pages={1175--1189},
  year={2022},
  publisher={Springer}
}

@article{pearce2024evaluation,
  title={Evaluation of an inverse method for quantifying spatially variable mechanics},
  author={Pearce, Daniel P and Witzenburg, Colleen M},
  journal={Journal of Biomechanical Engineering},
  volume={146},
  number={12},
  pages={121006},
  year={2024},
  publisher={American Society of Mechanical Engineers}
}

@article{tao2024experimental,
  title={Experimental approach for characterizing the nonlinear, time and temperature-dependent constitutive response of open-cell polyurethane foams},
  author={Tao, Jialiang and Sun, Xiangyu and Franck, Christian},
  journal={Strain},
  volume={60},
  number={6},
  pages={e12478},
  year={2024},
  publisher={Wiley Online Library}
}

@article{zhou2025vaginal,
  title={Vaginal biomechanical function in premenopausal and postmenopausal women with and without pelvic organ prolapse},
  author={Zhou, Qinhan and Li, Guang and Kiley, Jasmine X and Ogola, Benard and Danso, Elvis K and Baiamonte, Lyndsey Buckner and Desrosiers, Laurephile and Knoepp, Leise R and Lindsey, Sarah H and Florian-Rodriguez, Maria E and others},
  journal={Scientific Reports},
  volume={15},
  number={1},
  pages={27039},
  year={2025},
  publisher={Nature Publishing Group UK London}
}

@article{mcghee2023high,
  title={High-speed, full-field deformation measurements near inertial microcavitation bubbles inside viscoelastic hydrogels},
  author={McGhee, A and Yang, J and Bremer, EC and Xu, Z and Cramer III, HC and Estrada, JB and Henann, DL and Franck, CJEM},
  journal={Experimental Mechanics},
  volume={63},
  number={1},
  pages={63--78},
  year={2023},
  publisher={Springer}
}

@article{yang2026inertial,
  title={Inertial interface cavitation creates complex, flow-like structures within a soft solid},
  author={Yang, Jin and McGhee, Alexander and Tong, Zixiang and Radtke, Griffin and Rodriguez Jr, Mauro and Franck, Christian},
  journal={Experimental Mechanics},
  pages={1--19},
  year={2026},
  publisher={Springer}
}

@inproceedings{yang2025imac,
  title={Spatiotemporally-resolved Kinematic and Stress Measurements of Interfacial Cavitation in Soft Matter via {DIC}},
  author={ Yang, Jin and McGhee, Alexander and Tong, Zixiang and Bu, Lehu and Wang, Sicong and Radtke, Griffin and Rodriguez, Mauro and Franck, Christian },
  booktitle={Computer Vision \& Laser Vibrometry},
  volume={6},
  pages={1--12},
  year={2026},
}

@article{yang2021smart,
  title={Smart digital image correlation patterns via {3D} printing},
  author={Yang, J and Tao, JL and Franck, C},
  journal={Experimental Mechanics},
  volume={61},
  number={7},
  pages={1181--1191},
  year={2021},
  publisher={Springer}
}

@inproceedings{bu2024high,
  title={High-Speed, Full-Field Measurement of Large Deformations Near Needle-Induced Cavitation Bubbles Within Biological Soft Materials},
  author={Bu, Lehu and Hou, Zhao-Bang and Polidoro, Sophie and Yang, Jin},
  booktitle={SEM Annual Conference and Exposition on Experimental and Applied Mechanics},
  pages={115--125},
  year={2024},
  organization={Springer}
}

@article{harmal2023bioinspired,
  title={Bioinspired brick-and-mortar geopolymer composites with ultra-high toughness},
  author={Harmal, Anass and Khouchani, Oussama and El-Korchi, Tahar and Tao, Mingjiang and Walker, Harold W},
  journal={Cement and Concrete Composites},
  volume={137},
  pages={104944},
  year={2023},
  publisher={Elsevier}
}

@article{grubii2023impact,
  title={The Impact of Top-Layer Sliced Lamella Thickness and Core Type on Surface-Checking in Engineered Wood Flooring},
  author={Grub{\^\i}i, Victor and Johansson, Jimmy},
  journal={Forests},
  volume={14},
  number={11},
  pages={2250},
  year={2023},
  publisher={MDPI}
}

@article{grubii2023measurement,
  title={Measurement of surface-checking in sliced lamellae-based engineered wood flooring using digital image correlation},
  author={Grub{\^\i}i, Victor and Johansson, Jimmy and Dagbro, Ola},
  journal={European Journal of Wood and Wood Products},
  volume={81},
  number={6},
  pages={1427--1436},
  year={2023},
  publisher={Springer}
}

@article{grubii2023influence,
  title={{The Influence of Slicing Thickness on the Perpendicular to Grain Tensile Properties of Oak (Quercus robur L. and Quercus petraea L.) Lamellae}},
  author={Grub{\^\i}i, Victor and Johansson, Jimmy},
  journal={Applied Sciences},
  volume={13},
  number={22},
  pages={12254},
  year={2023},
  publisher={MDPI}
}

@article{falta2023mechanical,
  title={Mechanical properties of basalt: a study on compressive loading at different strain rates using {SHPB}},
  author={Falta, Jan and Kr{\v{c}}m{\'a}{\v{r}}ov{\'a}, Nela and F{\'\i}la, Tom{\'a}{\v{s}} and Vavro, Martin and Vavro, Leona},
  journal={Acta Polytechnica CTU Proceedings},
  volume={42},
  pages={17--21},
  year={2023}
}

@phdthesis{grubii2023quality,
  title={Quality Aspects of Sliced Oak Lamellae in Development of Engineered Wood Flooring},
  author={Grub{\^\i}i, Victor},
  year={2023},
  school={Linnaeus University Press}
}

@inproceedings{hafiz2025uniaxial,
  title={Uniaxial Tensile Stress Relaxation and Cracking Behavior of SHCCs Incorporating Blast Furnace Slag and Fly Ash},
  author={Hafiz Zadah, Faizudin and Luan, Yao},
  booktitle={Biennial RILEM Youth Symposium on Building Materials and Construction},
  pages={909--923},
  year={2025},
  organization={Springer}
}

@article{wu2023deformation,
  title={Deformation measurement within lithium-ion battery using sparse-view computed tomography and digital image correlation},
  author={Wu, Yapeng and Sun, Liang and Zhang, Xiangchun and Yang, Min and Tan, Dalong and Hai, Chao and Liu, Jing and Wang, Juntao},
  journal={Measurement Science and Technology},
  volume={34},
  number={2},
  pages={025402},
  year={2023},
  publisher={IOP Publishing}
}

@article{courant2024design,
  title={Design and characterization of an efficient multistable push-pull linear actuator using magnetic shape memory alloys},
  author={Courant, Robert and Maas, J{\"u}rgen},
  journal={IEEE Access},
  volume={12},
  pages={107855--107871},
  year={2024},
  publisher={IEEE}
}

@article{zhang2023pattern,
  title={Pattern evolution and modal decomposition of Faraday waves in a brimful cylinder},
  author={Zhang, Shimin and Borthwick, Alistair GL and Lin, Zhiliang},
  journal={Journal of Fluid Mechanics},
  volume={974},
  pages={A56},
  year={2023},
  publisher={Cambridge University Press}
}

@article{zhang2025transitional,
  title={Transitional response of double-mode Faraday waves in a brimful container},
  author={Zhang, Shimin and Lin, Zhiliang},
  journal={Physical Review Fluids},
  volume={10},
  number={3},
  pages={034003},
  year={2025},
  publisher={APS}
}

@inproceedings{leon2022influence,
  title={Influence of displacement gradients on laser speckle photography},
  author={Le{\'o}n, Schweickhardt and Andreas, Tausendfreund and Dirk, St{\"o}bener and Andreas, Fischer},
  booktitle={EPJ Web of Conferences},
  volume={266},
  pages={10020},
  year={2022},
  organization={EDP Sciences}
}

@article{zhang2025free,
  title={Free-surface topography measurements of fluid layers over a smoothly varying bed},
  author={Zhang, Shimin and Yang, Hao and Borthwick, Alistair GL and Lin, Zhiliang},
  journal={Experiments in Fluids},
  volume={66},
  number={11},
  pages={199},
  year={2025},
  publisher={Springer}
}

@article{tong2026raftcorr,
  title={{RAFTcorr: A} Deep Learning Digital Image Correlation Framework with Operating-Boundary Characterization},
  author={Tong, Zixiang and Bu, Lehu and Shi, Qihang and Du, Runtian and Yang, Jin},
  year={2026}
}

@article{yang2022serialtrack,
  title={{SerialTrack: ScalE and Rotation Invariant Augmented Lagrangian particle tracking}},
  author={Yang, Jin and Yin, Yue and Landauer, Alexander K and Buyukozturk, Selda and Zhang, Jing and Summey, Luke and McGhee, Alexander and Fu, Matt K and Dabiri, John O and Franck, Christian},
  journal={SoftwareX},
  volume={19},
  pages={101204},
  year={2022},
  publisher={Elsevier}
}

@article{YangR3DICnet,
author = {Jiashuai Yang and Kemao Qian and Lianpo Wang},
journal = {Optics Express},
number = {1},
pages = {907--921},
publisher = {Optica Publishing Group},
title = {{R$^3$-DICnet: an end-to-end recursive residual refinement DIC network for larger deformation measurement}},
volume = {32},
month = {Jan},
year = {2024},
doi = {10.1364/OE.505655},
}

@article{Ma21,
author = {Chaochen Ma and Qing Ren and Jian Zhao},
journal = {Optics Express},
number = {6},
pages = {9137--9156},
publisher = {Optica Publishing Group},
title = {Optical-numerical method based on a convolutional neural network for full-field subpixel displacement measurements},
volume = {29},
month = {Mar},
year = {2021},
url = {https://opg.optica.org/oe/abstract.cfm?URI=oe-29-6-9137},
doi = {10.1364/OE.417413},
}

@article{FENG2024108267,
title = {{Stereo-DICNet: An efficient and unified speckle matching network for stereo digital image correlation measurement}},
journal = {Optics and Lasers in Engineering},
volume = {179},
pages = {108267},
year = {2024},
issn = {0143-8166},
doi = {https://doi.org/10.1016/j.optlaseng.2024.108267},
url = {https://www.sciencedirect.com/science/article/pii/S014381662400246X},
author = {Yahong Feng and Lianpo Wang},
}

@article{kafka2024technique,
  title={A Technique for In-Situ Displacement and Strain Measurement with Laboratory-Scale {X}-Ray Computed Tomography},
  author={Kafka, OL and Landauer, AK and Benzing, JT and Moser, NH and Mansfield, E and Garboczi, EJ},
  journal={Experimental Techniques},
  volume={48},
  number={5},
  pages={935--935},
  year={2024},
  publisher={SPRINGER ONE NEW YORK PLAZA, SUITE 4600, NEW YORK, NY, UNITED STATES}
}

@article{YANG2025115908,
title = {Efficient and robust deformation measurement based on unsupervised learning},
journal = {Measurement},
volume = {242},
pages = {115908},
year = {2025},
issn = {0263-2241},
doi = {https://doi.org/10.1016/j.measurement.2024.115908},
url = {https://www.sciencedirect.com/science/article/pii/S0263224124017937},
author = {Jiashuai Yang and Yahong Feng and Lianpo Wang},
}

@article{gupta2024situ,
  title={In situ analysis of plastic flow near interfaces and free surfaces},
  author={Gupta, Deepika and Udupa, Anirudh and Viswanathan, Koushik},
  journal={Measurement Science and Technology},
  volume={35},
  number={4},
  pages={045601},
  year={2024},
  publisher={IOP Publishing}
}

@article{feng2026stereo,
  title={{Stereo-DICNet2: A Unified and Physics-Guided Speckle Matching Network for Three-Dimensional Deformation Measurement}},
  author={Feng, Y and Wang, L},
  journal={Experimental Mechanics},
  volume={66},
  number={2},
  pages={399--415},
  year={2026},
  publisher={Springer}
}

@inproceedings{Mangileva2024,
    author = { Mangileva, Daria},
    title = {The Optimal {MLP}-based Model for Displacement Field Measurement in {2D} Images and Its Application Perspective},
    booktitle = {CVIPPR '24: Proceedings of the 2024 2nd Asia Conference on Computer Vision, Image Processing and Pattern Recognition},
    volume={36},
    pages={1--8},
    year ={2024} 
}

@article{PUREZA2026102532,
title = {{DICLab2D: An open-source digital image correlation algorithm for Julia language}},
journal = {SoftwareX},
volume = {33},
pages = {102532},
year = {2026},
issn = {2352-7110},
doi = {https://doi.org/10.1016/j.softx.2026.102532},
url = {https://www.sciencedirect.com/science/article/pii/S2352711026000269},
author = {Dennis Quaresma Pureza and José Luis Vital {de Brito} and Guilherme Santana Alencar and Luís Augusto Conte Mendes Veloso},
}

@article{schweickhardt2023digital,
  author  = {Schweickhardt, Le{\'o}n and Tausendfreund, Andreas and St{\"o}bener, Dirk and Fischer, Andreas},
  title   = {Digital Speckle Photography in the Presence of Displacement Gradients},
  journal = {Journal of the European Optical Society-Rapid Publications},
  year    = {2023},
  volume  = {19},
  pages   = {16},
  doi     = {10.1051/jeos/2023012}
}

@article{venter2025sun,
  title={{SUN-DIC: A Python-based open-source software tool for Digital Image Correlation}},
  author={Venter, Gerhard and Neaves, Melody},
  journal={Advances in Engineering Software},
  volume={211},
  pages={104043},
  year={2025},
  publisher={Elsevier}
}

@article{wang2026interpretable,
  title={{Interpretable DIC measurement neural network based on IC-GN algorithm framework}},
  author={Wang, Lianpo and Zhang, Yantao},
  journal={Optics Express},
  volume={34},
  number={6},
  pages={9890--9905},
  year={2026},
  publisher={Optica Publishing Group}
}

@article{kopiika2024digital,
  title={Digital image correlation for assessment of bridges’ technical state and remaining resource},
  author={Kopiika, Nadiia and Blikharskyy, Yaroslav},
  journal={Structural Control and Health Monitoring},
  volume={2024},
  number={1},
  pages={1763285},
  year={2024},
  publisher={Wiley Online Library}
}

@article{naufal2024digital,
  title={{Digital image correlation technique for failure and crack propagation of fibre-reinforced polymer composites--A review}},
  author={Naufal, Muhammad Irfan and Wong, King Jye and Israr, Haris Ahmad and Nejad, Ali Farokhi and Rahimian Koloor, Seyed Saeid and Gan, Khong Wui and Faizi, Mohd Khairul and Siebert, Geralt},
  journal={Composites and Advanced Materials},
  volume={33},
  pages={26349833241253619},
  year={2024},
  publisher={SAGE Publications Sage UK: London, England}
}

@article{kibrete2025free,
  title={Free and Open-Source Python-Based Software Packages for Digital Image Correlation in Full-Field Displacement and Strain Measurements},
  author={Kibrete, Fasikaw and Woldemichael, Dereje Engida and Gebremedhen, Hailu Shimels and Batu, Temesgen},
  journal={Strain},
  volume={61},
  number={4},
  pages={e70012},
  year={2025},
  publisher={Wiley Online Library}
}

@article{zhu2026pf,
  title={{PF-DIC}: Phase field digital image correlation for integrated full-field displacement, strain, and damage measurements},
  author={Zhu, Dingxiang and Lu, Ye},
  journal={arXiv preprint arXiv:2606.16850},
  year={2026}
}

@article{jin2023recent,
  title={{Recent advances and applications of machine learning in experimental solid mechanics: A review}},
  author={Jin, Hanxun and Zhang, Enrui and Espinosa, Horacio D},
  journal={Applied Mechanics Reviews},
  volume={75},
  number={6},
  pages={061001},
  year={2023},
  publisher={American Society of Mechanical Engineers}
}

@article{wihardja2025constitutive,
  title={Constitutive relations from images},
  author={Wihardja, Adeline and Bhattacharya, Kaushik},
  journal={Journal of Applied Mechanics},
  volume={92},
  number={8},
  pages={081009},
  year={2025},
  publisher={American Society of Mechanical Engineers}
}

@article{pearce2022combining,
  title={Combining unique planar biaxial testing with full-field thickness and displacement measurement for spatial characterization of soft tissues},
  author={Pearce, Daniel and Nemcek, Mark and Witzenburg, Colleen},
  journal={Current Protocols},
  volume={2},
  number={7},
  pages={e493},
  year={2022},
  publisher={Wiley Online Library}
}

@article{you2022physics,
  title={A physics-guided neural operator learning approach to model biological tissues from digital image correlation measurements},
  author={You, Huaiqian and Zhang, Quinn and Ross, Colton J and Lee, Chung-Hao and Hsu, Ming-Chen and Yu, Yue},
  journal={Journal of Biomechanical Engineering},
  volume={144},
  number={12},
  pages={121012},
  year={2022},
  publisher={American Society of Mechanical Engineers}
}

@article{flaschel2021unsupervised,
  title={Unsupervised discovery of interpretable hyperelastic constitutive laws},
  author={Flaschel, Moritz and Kumar, Siddhant and De Lorenzis, Laura},
  journal={Computer Methods in Applied Mechanics and Engineering},
  volume={381},
  pages={113852},
  year={2021},
  publisher={Elsevier}
}

@article{kirchhoff2024inference,
  title={Inference of Heterogeneous Material Properties via Infinite-Dimensional Integrated DIC},
  author={Kirchhoff, Joseph and Luo, Dingcheng and O'Leary-Roseberry, Thomas and Ghattas, Omar},
  journal={arXiv preprint arXiv:2408.10217},
  year={2024}
}

@article{spencer2026full,
  title={Full-Field Calibration of Coupled Thermomechanical Material Models at Finite Strain},
  author={Spencer, L and Meador, William D and Tepole, Adrian Buganza and Granzow, Brian N and Yang, Jin and Rausch, Manuel K and Seidl, D Thomas and Fuhg, Jan N},
  journal={arXiv preprint arXiv:2606.05465},
  year={2026}
}

@article{kumar2025comparative,
  title={A comparative study of calibration techniques for finite strain elastoplasticity: Numerically-exact sensitivities for FEMU and VFM},
  author={Kumar, Sanjeev and Seidl, D Thomas and Granzow, Brian N and Yang, Jin and Fuhg, Jan Niklas},
  journal={Computer Methods in Applied Mechanics and Engineering},
  volume={444},
  pages={118159},
  year={2025},
  publisher={Elsevier}
}

@article{wei2026machine,
  title={Machine learning extraction of viscoelastic material properties from full-field deformation measurements},
  author={Wei, Congjie and Bu, Lehu and Yang, Jin and Wu, Chenglin},
  journal={Journal of the Mechanics and Physics of Solids},
  pages={106589},
  year={2026},
  publisher={Elsevier}
}

@article{pan2013fast,
  title={Fast, robust and accurate digital image correlation calculation without redundant computations},
  author={Pan, Bing and Li, Kai and Tong, Wei},
  journal={Experimental Mechanics},
  volume={53},
  number={7},
  pages={1277--1289},
  year={2013},
  publisher={Springer}
}

@article{landauer2018q,
  title={A q-factor-based digital image correlation algorithm ({qDIC}) for resolving finite deformations with degenerate speckle patterns},
  author={Landauer, Alexander K and Patel, M and Henann, DL and Franck, C},
  journal={Experimental Mechanics},
  volume={58},
  number={5},
  pages={815--830},
  year={2018},
  publisher={Springer}
}

@inproceedings{yang2018fast,
  title={Fast adaptive global digital image correlation},
  author={Yang, Jin and Bhattacharya, Kaushik},
  booktitle={Advancement of Optical Methods \& Digital Image Correlation in Experimental Mechanics, Volume 3: Proceedings of the 2018 Annual Conference on Experimental and Applied Mechanics},
  pages={69--73},
  year={2018},
  organization={Springer}
}

@article{chen2017active,
  title={Active slip system identification in polycrystalline metals by digital image correlation ({DIC})},
  author={Chen, Z and Daly, SH},
  journal={Experimental Mechanics},
  volume={57},
  number={1},
  pages={115--127},
  year={2017},
  publisher={Springer}
}

@article{rubino2019full,
  title={Full-field ultrahigh-speed quantification of dynamic shear ruptures using digital image correlation},
  author={Rubino, V and Rosakis, AJ and Lapusta, N},
  journal={Experimental Mechanics},
  volume={59},
  number={5},
  pages={551--582},
  year={2019},
  publisher={Springer}
}

@article{gupta2024estimation,
  title={Estimation of surface and interface strains in deformation processing using an ensemble averaged digital image correlation method},
  author={Gupta, Deepika and Viswanathan, Koushik},
  journal={Journal of Manufacturing Processes},
  volume={120},
  pages={86--95},
  year={2024},
  publisher={Elsevier}
}

@article{baker2004lucas,
  title={{Lucas-Kanade 20 years on: A unifying framework}},
  author={Baker, Simon and Matthews, Iain},
  journal={International Journal of Computer Vision},
  volume={56},
  number={3},
  pages={221--255},
  year={2004},
  publisher={Springer}
}

@article{boyd2011distributed,
	title = {Distributed {Optimization} and {Statistical} {Learning} via the {Alternating} {Direction} {Method} of {Multipliers}},
	volume = {3},
	issn = {1935-8237},
	url = {http://dx.doi.org/10.1561/2200000016},
	number = {1},
	journal = {Foundations and Trends in Machine Learning},
	author = {Boyd, Stephen and Parikh, Neal and Chu, Eric and Peleato, Borja and Eckstein, Jonathan},
	year = {2011},
	pages = {1--122},
}

@article{yang2020augmented,
  title={{Augmented Lagrangian Digital Volume Correlation (ALDVC)}},
  author={Yang, J and Hazlett, L and Landauer, A. K. and Franck, C},
  journal={Experimental Mechanics},
  volume={60},
  number={9},
  pages={1205--1223},
  year={2020},
  url={https://doi.org/10.1007/s11340-020-00607-3}
}

@article{yang2019augmented,
  title={Augmented {L}agrangian digital image correlation},
  author={Yang, J and Bhattacharya, K},
  journal={Experimental Mechanics},
  volume={59},
  number={2},
  pages={187--205},
  year={2019},
  url={https://doi.org/10.1007/s11340-018-00457-0}
}

@article{dicchallenge2exme,
title={{DIC Challenge 2.0: Developing Images and Guidelines for Evaluating Accuracy and Resolution of 2D Analyses}},
author={Reu, P L and Blaysat, Benoit and And{\`o}, Edward and Bhattacharya, Kaushik and Couture, Cyrille and Couty, Vincent and Deb, Debasis and Fayad, S S and Iadicola, M A and Jaminion, St{\'e}phanie and others},
journal={Experimental Mechanics},
pages={1--16},
year={2022},
publisher={Springer},
url = {https://doi.org/10.1007/s11340-021-00806-6},
}

@article{yang2022staq,
author={J Yang and V Rubino and Z Ma and JL Tao and Y Yin and A McGhee and WX Pan and C Franck},
title={{SpatioTemporally Adaptive Quadtree mesh (STAQ) Digital Image Correlation for resolving large deformations around complex geometries and discontinuities}},
journal={Experimental Mechanics},
year={2022}
}

@article{tong2025stereo3d,
  title   = {{3D Stereo Adaptive Mesh Augmented Lagrangian Digital Image Correlation}},
  author  = {Tong, Zixiang and Frolkin, Danila and Shi, Hongyang and LaRue Trace and Rausch, Manuel K. and Zhang, Yujie and Knight, Virginia and Yang, Jin},
  journal = {Experimental Mechanics},
  year    = {2025},
  volume = {65},
  pages = {1387-1411},
  doi     = {10.1007/s11340-025-01225-7},
  url     = {https://doi.org/10.1007/s11340-025-01225-7}
}

@book{idics_gpg_2018,
  title     = {A Good Practices Guide for Digital Image Correlation},
  editor    = {Jones, Elizabeth M. C. and Iadicola, Mark A.},
  publisher = {International Digital Image Correlation Society (iDICs)},
  year      = {2018},
  edition   = {1st},
  doi       = {10.32720/idics/gpg.ed1},
  url       = {https://doi.org/10.32720/idics/gpg.ed1}
}

@book{idics_gpg_2025,
  title     = {A Good Practices Guide for Digital Image Correlation},
  editor    = {Jones, Elizabeth M. C. and Iadicola, Mark A. and Yang, Jin and Tong, Zixiang and others},
  publisher = {International Digital Image Correlation Society (iDICs)},
  year      = {2025},
  edition   = {2nd},
  note      = {2nd edition; co-edited by authors of the present work}
}

@article{blaber2015ncorr,
  title   = {{Ncorr: Open-Source 2D Digital Image Correlation Matlab Software}},
  author  = {Blaber, J. and Adair, B. and Antoniou, A.},
  journal = {Experimental Mechanics},
  volume  = {55},
  number  = {6},
  pages   = {1105--1122},
  year    = {2015},
  doi     = {10.1007/s11340-015-0009-1},
  url     = {https://doi.org/10.1007/s11340-015-0009-1}
}

@misc{turner2015dice,
  author       = {Turner, D. Z. and Lehoucq, R. B. and Reu, P. L. and Garavito-Camargo, N. and others},
  title        = {{DICe: Digital Image Correlation Engine}},
  howpublished = {Sandia National Laboratories},
  year         = {2015},
  url          = {https://github.com/dicengine/dice}
}

@article{olufsen2020mudic,
  title   = {{\textmu}DIC: An open-source toolkit for digital image correlation},
  author  = {Olufsen, S. N. and Andersen, M. E. and F{\o}rre, E.},
  journal = {SoftwareX},
  volume  = {11},
  pages   = {100391},
  year    = {2020},
  doi     = {10.1016/j.softx.2019.100391},
  url     = {https://doi.org/10.1016/j.softx.2019.100391}
}

@misc{vic2d_software,
  title        = {{VIC-2D Digital Image Correlation Software}},
  author       = {{Correlated Solutions, Inc.}},
  howpublished = {\url{https://www.correlatedsolutions.com/vic-2d/}},
  year         = {2024},
  note         = {Commercial software; accessed 2026-05-13}
}

@misc{DaVis_software,
  title        = {{DaVis Software}},
  author       = {{LaVision}},
  howpublished = {\url{https://www.lavision.de/en/products/davis-software/}},
  year         = {},
  note         = {Commercial software; accessed 2026-07-15}
}

@misc{EikoSim_software,
  title        = {{EikoSim Software}},
  author       = {{EikoSim}},
  howpublished = {\url{https://eikosim.com/}},
  year         = {},
  note         = {Commercial software; accessed 2026-07-15}
}

@misc{matchid_software,
  title        = {{MatchID 2D Digital Image Correlation Software}},
  author       = {{MatchID NV}},
  howpublished = {\url{https://www.matchid.eu/}},
  year         = {2024},
  note         = {Commercial software; accessed 2026-05-13}
}

@misc{gom_aramis_software,
  title        = {{ARAMIS 3D Optical Strain and Displacement Analysis}},
  author       = {{ZEISS / GOM}},
  howpublished = {\url{https://www.gom.com/en/products/zeiss-aramis}},
  year         = {2024},
  note         = {Commercial software; accessed 2026-05-13}
}

@inproceedings{lam2015numba,
  title     = {{Numba}: A {LLVM}-based {Python} {JIT} Compiler},
  author    = {Lam, Siu Kwan and Pitrou, Antoine and Seibert, Stanley},
  booktitle = {Proceedings of the Second Workshop on the LLVM Compiler Infrastructure in HPC},
  pages     = {1--6},
  year      = {2015},
  doi       = {10.1145/2833157.2833162}
}

@misc{pyaldic_zenodo,
  author       = {Tong, Zixiang and Yang, Jin},
  title        = {{pyALDIC: Augmented Lagrangian Digital Image Correlation in Python}},
  year         = {2026},
  howpublished = {Zenodo software archive, version v0.6.0},
  doi          = {10.5281/zenodo.19521061},
  note         = {\url{https://doi.org/10.5281/zenodo.19521061}}
}

@article{opencorr,
  author  = {Jiang, Zhenyu},
  title   = {{OpenCorr}: An open source library for research and development of digital image correlation},
  journal = {Optics and Lasers in Engineering},
  volume  = {165},
  pages   = {107566},
  year    = {2023},
  doi     = {10.1016/j.optlaseng.2023.107566}
}

@article{icorrvision,
  author  = {de Deus Filho, Jo{\~a}o Carlos Andrade and Nunes, Luiz Carlos da Silva and Xavier, Jos{\'e} Manuel Cardoso},
  title   = {{iCorrVision-2D}: An integrated python-based open-source Digital Image Correlation software for in-plane measurements ({Part}~1)},
  journal = {SoftwareX},
  volume  = {19},
  pages   = {101131},
  year    = {2022},
  doi     = {10.1016/j.softx.2022.101131}
}

@article{yang2021fast,
  author  = {Yang, J. and Bhattacharya, K.},
  title   = {Fast Adaptive Mesh Augmented {L}agrangian Digital Image Correlation},
  journal = {Experimental Mechanics},
  volume  = {61},
  number  = {4},
  pages   = {719--735},
  year    = {2021},
  doi     = {10.1007/s11340-021-00695-9}
}

@article{poissant2010novel,
  author  = {Poissant, J. and Barthelat, F.},
  title   = {A Novel ``Subset Splitting'' Procedure for Digital Image Correlation on Discontinuous Displacement Fields},
  journal = {Experimental Mechanics},
  volume  = {50},
  number  = {3},
  pages   = {353--364},
  year    = {2010},
  doi     = {10.1007/s11340-009-9220-2}
}

@article{pan2018review,
  author  = {Pan, Bing},
  title   = {Digital image correlation for surface deformation measurement: historical developments, recent advances and future goals},
  journal = {Measurement Science and Technology},
  volume  = {29},
  number  = {8},
  pages   = {082001},
  year    = {2018},
  doi     = {10.1088/1361-6501/aac55b}
}

@book{sutton2009,
  author    = {Sutton, Michael A. and Orteu, Jean-Jos{\'e} and Schreier, Hubert W.},
  title     = {Image Correlation for Shape, Motion and Deformation Measurements: Basic Concepts, Theory and Applications},
  publisher = {Springer},
  address   = {New York},
  year      = {2009},
  doi       = {10.1007/978-0-387-78747-3}
}

@article{leu2026machine,
  title={Machine Learning-Aided Spatial Adaptation for Improved Digital Image Correlation Analysis of Complex Geometries},
  author={Leu, Jeffrey and Tong, Zixiang and Doty, Andrew and Tsimpoukis, Solon and Deng, Bolei and Yang, Jin},
  journal={Strain},
  volume={62},
  number={1},
  pages={e70022},
  year={2026},
  publisher={Wiley Online Library}
}

@article{yang2019combining,
  title={Combining image compression with digital image correlation},
  author={Yang, Jin and Bhattacharya, Kaushik},
  journal={Experimental Mechanics},
  volume={59},
  number={5},
  pages={629--642},
  year={2019},
  publisher={Springer}
}

@misc{rethore2018ufreckles,
  author       = {R{\'e}thor{\'e}, Julien},
  title        = {{UFreckles}},
  year         = {2018},
  publisher    = {Zenodo},
  doi          = {10.5281/zenodo.1433776},
  note         = {Open-source MATLAB global (FE-based) DIC/DVC with GUI},
  url          = {https://doi.org/10.5281/zenodo.1433776}
}

@misc{yadics,
  title        = {{YaDICs}: Yet Another Digital Image Correlation software},
  howpublished = {\url{http://yadics.univ-lille1.fr/}},
  note         = {Open-source C\texttt{++} DIC (University of Lille); accessed 2026-07-08}
}

@misc{predic,
  title        = {{PReDIC}: {Python} Digital Image Correlation},
  howpublished = {\url{https://github.com/texM/PReDIC}},
  note         = {Open-source Python subset-based DIC; accessed 2026-07-08}
}

@article{li7088402opencorr,
  title={OpenCorr GUI: A free software for accurate and robust deformation measurement of digital image correlation and digital volume correlation},
  author={Li, Rui and He, Kai and Ren, Haoqiang and Wang, Hongli and Zhang, Yajing and Peng, Chang and Zhou, Yifei and Sun, Taolin and Tang, Liqun and Jiang, Zhenyu},
  journal={Available at SSRN 7088402}
}

\end{document}